# Smart mechanically tunable surfaces with shape memory behavior and wetting-programmable topography

*Gissela Constante, Indra Apsite, Paul Auerbach, Sebastian Aland, Dennis Schönfeld, Thorsten Pretsch, Pavel Milkin and Leonid Ionov*[*]*

G. Constante, I. Apsite, P. Milkin, L. Ionov
Faculty of Engineering Sciences, University of Bayreuth, Ludwig Thoma Str. 36A, 95447 Bayreuth, Germany
E-mail: leonid.ionov@uni-bayreuth.de

P. Auerbach, S. Aland

Fakultät Mathematik und Informatik, TU Freiberg, 09599 Freiberg, Germany

D. Schönfeld, T. Pretsch

Fraunhofer Institute for Applied Polymer Research IAP, Geiselbergstr. 69, 14476 Postdam, Germany

L. Ionov

Bavarian Polymer Institute, University of Bayreuth, Bayreuth, Germany.

This contribution reports for the first time on fabrication and investigation of wetting properties of structured surfaces containing lamellae with an exceptionally high aspect ratio - height/width ratio demonstrated of 57:1. The lamellar surface was made using a polymer with tunable mechanical properties, and shape-memory behavior. It was found that wetting properties of such structured surfaces depend on temperature and thermal treatment history – lamellae are wetted easier at elevated temperature or after cooling to room temperature when the polymer is soft because of the easier deformability of lamellae. The shape of lamellae deformed by droplets can be temporarily fixed at low temperature and remains fixed upon heating to room temperature. Heating above the transition temperature of the shape-memory polymer restores the orginal shape. The high aspect ratio allows tuning of geometry not only manually, as it is done in most works reported previously, but can also be made by a liquid droplet and is controlled by temperature. The liquid in combination with thermoresponsive topography present a new kind of wetting behavior. Moreover, the mechanical properties can be controlled – the polymer can either be hard or soft at room temperature depending on thermal pre-history. This behavior opens new opportunities for the design of novel smart elements for microfluidic devices such as smart valves, whose state and behavior can be switched by thermal stimuli: valves can or cannot be opened, are able to close or can be fixed in an open or closed states.

## 1. Introduction

Surface wettability and structural design possibilities for its manipulation became a subject of intense research over the last decades because of their importance for material design,[1, 2] control of biological processes,[3, 4] and even for the technological industry.[5, 6] For example, highly hydrophilic and superhydrophobic materials can be useful for the development of self-cleaning coatings that are used to reduce corrosion, weathering, and erosion[7-9] or to prevent the accumulation of marine organisms on boats.[10-13] Various kinds of wetting behavior are also often observed in nature and are used by living organisms for adaptation and to increase their survival rate. For example, *Nelumbo,* the lotus plant, possesses a superhydrophobic surface on their leaves, where the presence of hierarchical micropapillae with a combination of epicuticular wax crystals on top produces an anisotropic surface. This arrangement allows rolling and sliding of the water droplets, which collect dust and dirt particles from the leaf surface allowing self-cleaning without energy consumption.[14] Another example is rose petals, which possess microscale papillae that are partially covered with nanoscale striae[15] facilitating the pinning of water to help them stay hydrated.[15, 16] In turn, rice leaves use aligned anisotropic topographic lamellae to produce a bidirectional wettability.[17] Their surface is covered with micro- and nanoscale structures that form a superhydrophobic surface for self-cleaning and water repellency.[18] Also, animals such as geckos and desert beetles have developed surfaces with specific wettability to overcome harsh environmental conditions. Geckos have a specific arrangement on the skin at their feet to improve adherence by varying the surface through a transition between Wenzel and Cassie state[19]. Desert beetles have a bumpy back surface for collecting water from fogs.[20]

Active control (switching) or passive adaptation of wetting properties can allow new applications of engineering materials and are beneficial for applications in biotechnology[21] like cell capture/release[22] and bio-detection,[23] microfluidics,[24] textiles,[25] fabrication of sensors,[26] and others. For example, Minko et. al. developed self-adaptative surfaces (SAS) based on polystyrene/poly(vinyl pyridine) brushes[27]. The design of smart surfaces with switchable/adaptive wetting is done primarily through two approaches. The first assumes the change in surface chemistry due to the incorporation of sensitive molecules that can respond to a certain stimulus, e.g. pH,[28],[29],[30] light,[31] temperature[32, 33], and others. The second approach assumes the switching of the surface topography. Surfaces with specific topography are usually fabricated from materials with tunable mechanical properties, which are exposed to external deformation afterwards like pressing, stretching, or bending.[28, 34] Fabrication of surfaces with switchable topography requires both: the development of materials with tailorable properties and, methods to fabricate structured surfaces. Examples of materials used for the fabrication of switchable topography surfaces are shape-memory polymers,[16, 35] hydrogels,[28] and liquid crystalline elastomers.[36-38]. Diverse methods including photolithography,[21, 39-41] stereolithography,[42] soft-lithography,[43] etching,[41, 44] and self-assembly[45, 46] are widely used for the preparation of such surfaces.

So far, smart surfaces with switchable topography have been prepared using the replica-molding method, by photolithography, stereolithography, and soft-lithography allowing the formation of micro-bumps with a height of a few microns to almost 100 µm.[28, 39-41, 47, 48] The typical aspect ratio (height / width ratio) of such structures is relatively low: 0.25:1,[47] 1:1,[41] 2:1,[39] 3:1.[28] Such micro-bump shape-memory structures provide superhydrophobic behavior in their initial state. Mechanical programming (deformation) changes their wetting properties and the surfaces behave like rice leaves: water droplets are pinned.[49] Similarly, structures mimicking the behavior of rice leaf surfaces can be made using electrospinning, although this technique provides uneven deposition of fibers leading to a high variation in the fiber mat thickness and distance. With this method, the height of features remains low (~ 300 nm) and the distance between the fiber arrays varies between 0.25 and 1 µm.[50] Although these smart surfaces show interesting programmable hydrophilic and hydrophobic properties, their height is limited to a few micrometers and the aspect ratio is not higher than 3:1. Due to the low aspect ratio, the features cannot be substantially deformed by droplets because of surface tension (the deflection amplitude of a beam decreases with its length at the same applied force). Therefore, manual deformation is always required to program the shape of the surface feature that substantially restricts the applicability of such surfaces.

In this article, we report for the first time on the fabrication of polymeric lamellar structures with tunable mechanical properties and shape-memory behavior, which can or cannot be deformed by water droplets depending of conditions, and investigate their surface-specific wetting properties. The individual structures were characterized by high aspect ratios, nominally by a height/width ratio up to 57:1, and they were studied with regard to their shape-memory behavior and tunable thermomechanical properties. The high aspect ratio allows tuning of geometry not only manually, as it is done in most works reported previously, but can also be realized by placing a liquid droplet on top of the structure and is controlled by temperature. In other words, the effect of a liquid in combination with thermoresponsive topography and wetting properties was studied. This approach opens new technological opportunities for the design of smart elements for microfluidic devices such as smart valves, which (i) cannot be opened at low temperature, (ii) can be opened by liquid at high temperature, (iii) can be left in an open state after cooling when liquid is applied and (iv) close at elevated temperature if no liquid is applied. In this way, an approach is pursued in which so-called "If-Then-Else" relationships apply. Such approaches of self-sufficiently reacting material systems are currently being intensively researched in connection with programmable materials[51-56].

In contrast to previously reported surface patterning techniques, in this work we have used melt-electrowriting (MEW) to allow the fabrication of high aspect ratio features and expand the possibilities of the available materials. MEW is a novel, solvent-free additive manufacturing technique that relies on a combination of 3D printing and electrospinning that enables programmed deposition of polymeric microfibers with precision around 1 μm.[57] MEW allows the deposition of different classes of melt-processable polymers such as polyesters[58, 59] and thermoplastic elastomers.[60, 61] The main advantage of MEW of thermoplastic polymer materials is the possibility of depositing very thin continuous fibers (1 – 100 μm) one on top of the other in a programmable way[57, 60, 62, 63] to produce a high aspect ratio of the "walls" or lamellae (aspect ratio can be more than 100:1). As material technology for the surface fabrication, we have used a thermoplastic polyurethane (TPU) with shape-memory behavior, which offers very interesting combinations of properties such as thermally adaptive mechanical properties and the ability to recover shape on heating.[64, 65] As a matter of fact, the high aspect ratio of surface features, which is enabled by MEW, allows their deformation by water droplets. In combination with the switchability of mechanical properties together with pronounced shape-memory properties, it allows a reversible programming of the topology/morphology and wetting state of the surface affected by temperature and the presence of water droplets.

## 2. Results and Discussion

As material basis for the present work we synthesized a phase segregated thermoplastic poly(ester-urethane) (PEU). The hard segments of the PEU were built up by 4,4´-diphenylmethane diisocyanate (MDI) and 1,4-butanediol (BD), whereas poly(1,4-butylene adipate) (PBA) was used as soft segment. Since the weigth percentage of the PBA phase was approximately 75%, the abbreviation TPU PBA-75 will be used hereafter. The physical properties of such thermoplastic polyurethanes are known to depend, among others, on the molecular weight of the soft segment, the molar ratio between hard and soft segments, and the polymerization process.[66-70]

The DSC thermogram of the synthesized PEU shows a clear peak associated with the melting of the soft segment including an onset temperature of 28 °C and a maximum at 40 °C (**Figure 1a**).

When cooling, the onset of the polymer's crystallization transition is at 10 °C and its maximum is at 5 °C. The degree of crystallinity ($\chi_c$), as calculated from the melting enthalpy of $\Delta H_m$ = -22.18 J g$^{-1}$, is $\chi_c$ = 21.9 %. For pure PBA the melting enthalpy is 135 J g$^{-1}$ ($\Delta H_m$ = 135 J g$^{-1}$, $\chi_c$ = 100 %).[71] Discrepancies between melting and crystallization temperatures of pure PBA and PBA-based TPU can be explained by the fact that in case of the TPU the crystallization of soft segments is prevented in parts by the presence of hard polyurethane segments. In other words, TPU PBA-75 is a classical physically crosslinked shape-memory polymer with a crystallizable soft segment. Interestingly, due to the soft segment phase transitions, the PBA phase can be both amorphous or semicrystalline at room temperature for at least a certain time, depending on its previous exposure to high or low temperatures.

The mechanical properties of the polymer were characterized by small-amplitude dynamic mechanical analysis (DMA) (**Figure 1**b) and cyclic stretching with different amplitudes at a constant rate (**Figure 1**c). For the DMA experiment, the polymer was exposed to 4 °C overnight in order to ensure the distinct crystallization of the soft segment, and then it was heated to room temperature. Since large parts of the PBA crystals melt at ca 40 °C, the soft segment is expected to be in a semicrystalline state at room temperature. Indeed, we observed that the pre-cooled polymer is initially hard at room temperature (E´ ≈ 55 MPa), and heating above the melting peak temperature (ca. 40 °C) makes the polymer softer (E´ ≈ 8 MPa). The same polymer, which was not exposed to 4 °C, but was heated to … °C and the cooled to room temperature where it was kept for … h, does not show a change of elastic modulus upon heating from room temperature to 80 °C meaning that soft segments remained mainly amorphous at room temperature and require lower temperatures to crystallize, showing a dependency on the thermal pretreatment. We performed cyclical mechanical tests at two different temperatures: 50 °C (slightly above the melting temperature of the PBA phase) and 70 °C (well above the melting temperature of the PBA phase) to investigate the mechanical properties (modulus and reversibility of deformation) at different amplitudes of deformation (**Figure 1c**). It was found that an increase in temperature from 50 °C to 70 °C results in a decrease of elastic modulus from 50 ± 5 MPa down to 9 ± 1 MPa (**Table S1**). Moreover, it was found that the mechanical reversibility and resilience increase upon heating to 70 °C (**Table 1**). These results fit with the deformation behavior of similar shape-memory poly(ester urethanes) described previously.[72, 73] Thus, the polymer is hard at room temperature, when it is preliminary exposed to 4 °C, and soft and elastic (deformation is reversible) at 70 °C. In shape memory studies, lamellae were bent at 70 °C and the imposed shape was fixed by cooling to 4 °C. Upon heating to 70 °C the lamellae recovered their original shape. Thus, such structures demonstrate promising surface-related shape memory properties. As already mentioned, an adjustable elastic modulus and elastic/plastic behavior upon heating and cooling are further characteristics, including that the polymer can be soft or hard at room temperature depending on thermal pretreatment. In detail, when heating the polymer to … °C and cooling it to room temperature where it was stored for … h, no PBA melting signal could be detected in the DSC thermogram within 48 h when reheating the polymer to … °C. We expect that PBA crystallization was temporarily hindered under these conditions due to the underlying kinetics.

Previous studies [70] showed that the morphological state of such polymers remains unchanged for approximately 48 hours. The specific temperatures, at which the polymer switches to soft (above room temperature) and hard (below room temperature but above water freezing point) states, as well as the specific temperature range (around room temperature), within it can be either in soft or hard, solely depend on thermal treatment. This make the shape memory polymer a highly interesting candidate for fabricating devices with tunable and programmable thermomechanical properties.

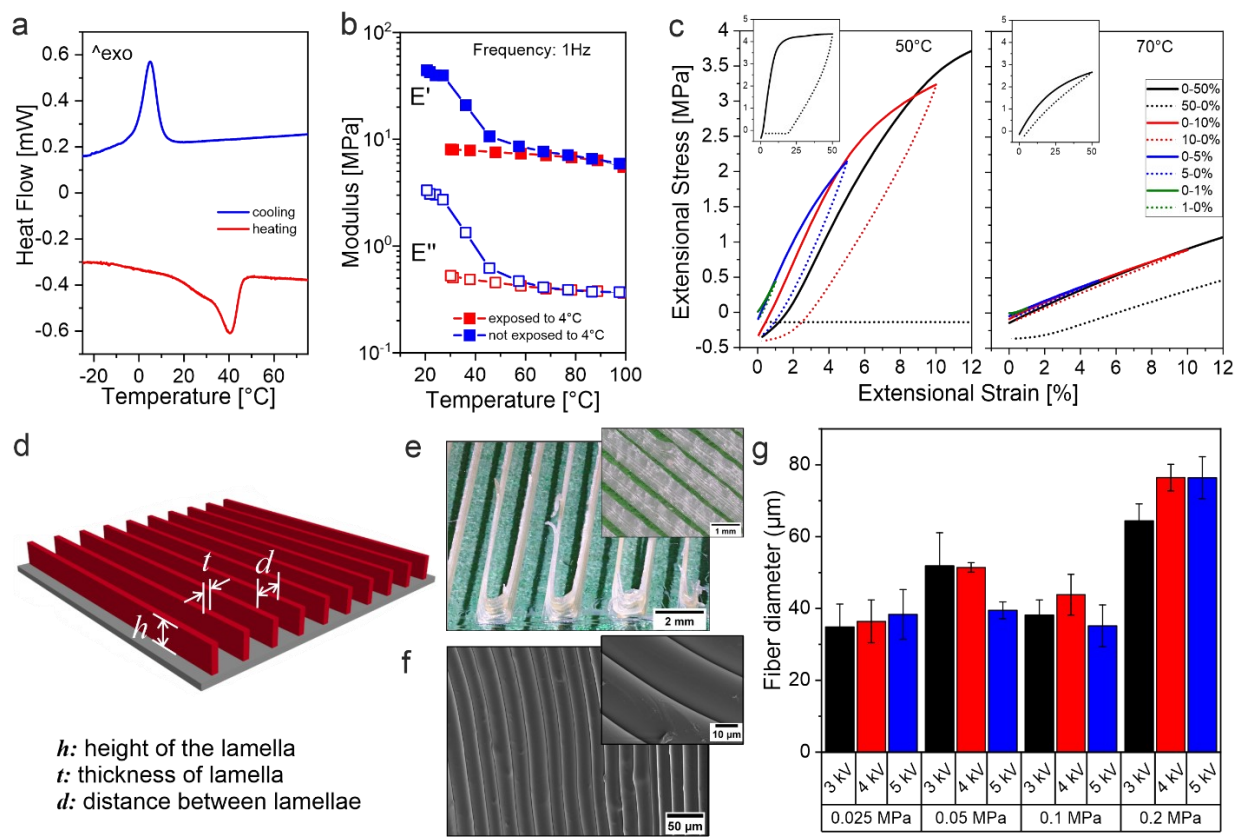

**Figure 1.** Thermal and mechanical properties of TPU PBA-25 and morphology of structured surfaces fabricated from TPU PBA-75 using melt electrowriting. (a) – DSC; (b) – DMA of polymer, which was exposed to 4°C (blue) and which was not exposed to 4°C (red); (c) – results of cyclical stretching; (d) schematics and (e, f) microscopy images (optical microscopy) and (SEM) of the morphology of fabricated surfaces; (g) dependence of fiber diameter on conditions of melt electrowriting.

**Table 1** Resilience and recovery of the TPU PBA-75 measured at the melting point of the soft segment and temperature of deformation used for the shape-memory studies.

| Temperature, [°C] | Max. Strain, [%] | Resilience, [%] | Recovery, [%] |
|---|---|---|---|
| 50 | 5 | 60.38 | 77.75 |
| 70 | 5 | 86.68 | 90.78 |

We fabricated surfaces from TPU PBA-75 with patterned topography. A design of parallel lamellae with different height, which is determined by the number of layers, and the distance between them was made by melt electrowriting (**Figure 1d/e/f**) on a glass slide substrate coated by a thin layer of polylactic acid (PLA). High aspect ratio of lamellae was achieved with the stacking of fibers (height/width ratio up to 57:1). The role of PLA was to increase the contact angle of water on the glass slide (< 40°) to 70° to prevent water spreading, and to improve fiber adhesion to the substrate. We studied the effects of MEW conditions on the diameter of the fiber and the uniformity of the lamellae (**Figure 1g**). Pressure and voltage were found to be crucial to define the smoothness, deposition, and diameter of the fibers deposited by MEW. In particular, an increase in pressure results in an increase in the diameter of deposited fibers, because higher pressure increases flow rate through the nozzle. The average diameter of the fiber is $37 \pm 2$ μm and $72 \pm 7$ μm, at 0.025 MPa and 0.2 MPa, respectively. The precision of the deposition of the fibers was found to be highly dependent on the voltage applied during the spinning. Very high voltages (5 kV) result in the formation of buckled fibers (sinusoidal shape, **Figure S2**) that is due to the high ratio between the fiber stretching and fiber deposition rate. Finally, 3 kV pressures between 0.025 and 0.1 MPa were chosen for the fabrication of small-diameter fibers (around 30 μm). Using these conditions, we fabricated lamellar surfaces with different heights (500 and 1700 μm) and distances (500 and 150 μm). The thickness of lamellae was kept at an average of 30 μm. The roughness coefficient was calculated for the topographical surface using the equation: $r = \frac{total\ area}{projected\ area}$ (**Figure S3**). Values from 1.65 to 7.36 were obtained by varying the distance *(d)* and the height *(h)* of the lamellae.

Adjacently, we investigated how droplets fill the space between the lamellae upon their advancing. The shape of the droplets, deposited in the grove between the lamellae, is spherical right after deposition and gets elongated along the lamellae during advancing (**Figure 2**) i.e., the wetting of structured surfaces is anisotropic. We modeled the behavior of a droplet between two stiff undeformable lamellae (the thickness of the lamellae was set to 100 μm). Numerical simulations were performed by a finite difference simulation of the Navier-Stokes equations[74] using a highly performant graphic processing unit (GPU) parallelization.[75] To regularize the contact line singularity the droplet and ambient air were modeled by a phase field function. The droplet was initially prescribed as a ball of radius 781 μm. The rigid lamellae were modeled by another stationary phase field and the contact angle of 72 degrees was imposed by the diffuse domain method.[76] The simulation shows how the droplet adapts its shape to adjust the contact angle and the dynamics of which are governed by the interplay of viscous and capillary forces. Thus, the modeling confirmed that the droplet must be initially rounded and shall get elongated during its development.

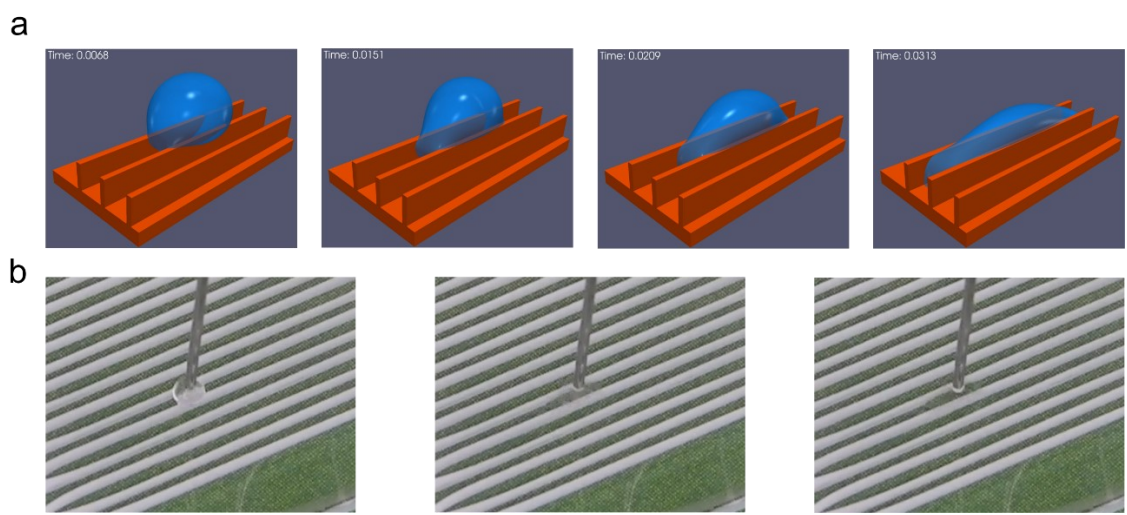

**Figure 2.** a) Simulation and b) experimental results of water behavior on a thick and hard lamellae structure.

In accordance to the results of thermomechanical analysis, the lamellae showed different behavior at room temperature when they were exposed to different thermal treatments. The lamellae expressed "softness mechanical memory" when the soft segment was molten previously at higher temperatures (>70°C). While the "hardness mechanical memory" was observed when the soft segment of the polymer remains crystalline at 4°C. The memory of mechanical properties was evidenced by quasi-rheological studies performed on the lamellae surfaces – the mechanical properties of lamellae deposited on the surface were tested by an oscillating plate in the setup used for rheological measurements. For this, the structured surfaces were exposed to high (70 °C) or low (4 °C) temperature for five minutes. Then the temperature was reduced/increased to 20 °C (RT) gradually. G´and G´´ were measured (**Figure 3a**). It was found that the same lamellae were hard and soft at room temperature regardless of whether they were previously cooled down or heated, respectively. The results have been treated qualitatively – the obtained value does not correspond to the shear modulus of the bulk polymer. However, the trend is the same - the surface formed by the lamellae is either hard or soft at room temperature, depending on the thermal pretreatment. Thus, the mechanical properties and wetting behavior are temporarily adjustable.

The different mechanical behaviors of the lamellae according to different thermal histories are depicted in (**Figure 3b, c**). At room temperature, a water droplet was deposited between two lamellae, which were exposed before to low (**Figure 3c**) and high (**Figure 3b**) temperatures. We observed that water droplets are able to deform the lamellae when they are soft (**Figure 3b**) and not able to deform them when they are hard (**Figure 3c**). Thus, the experiment clearly indicates different mechanical properties of lamellae when they were thermally treated, and on the fact that the lamellae can or cannot be deformed by capillary forces depending on their mechanical stiffness.

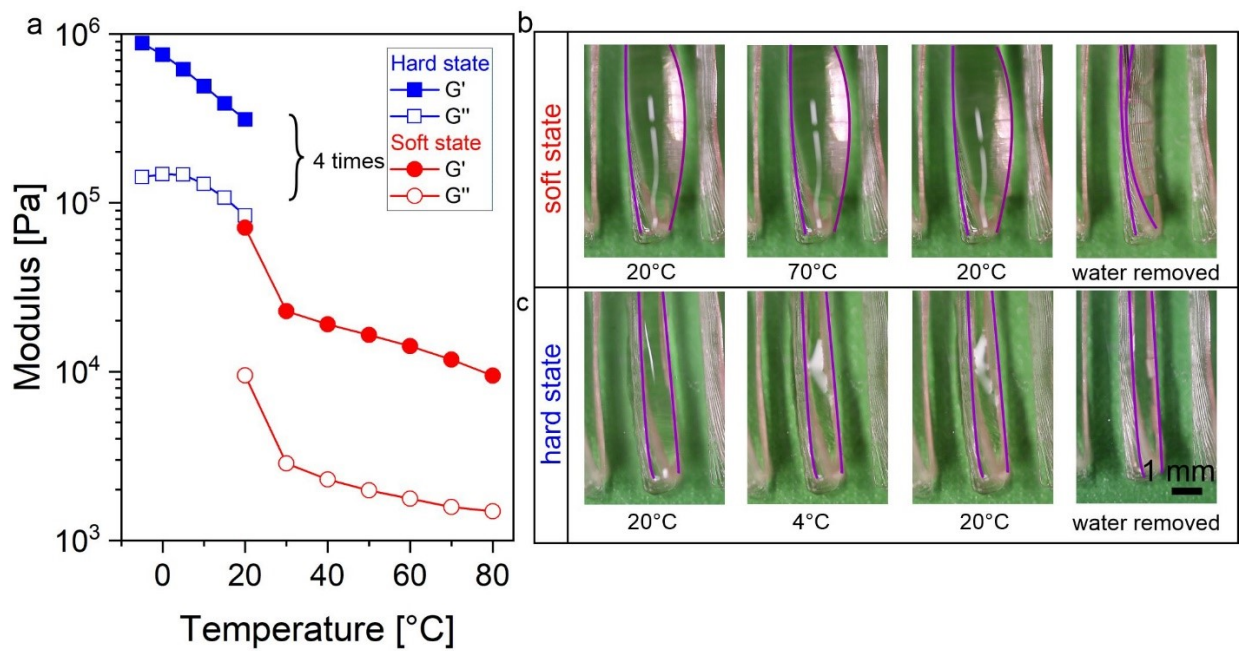

**Figure 3.** Thermomechanical studies of TPU PBA-75, a) rheological studies of hard lamellae (thermal pretreatment: cooling to -5 °C and heating to 20 °C) and soft lamellae (thermal pretreatment heating to 80 °C and cooling to 20 °C), b) images of the behavior analysis of the lamellae when it has been exposed to 70 °C and c) exposed to 4 °C.

The wetting behavior of the topographical surface was analyzed through advancing and receding volume measurements (**Figure 4**) when the material was exposed to 70 °C, further in the text referred to as "soft" and, when the material was exposed to 4 °C further in the text referred to as "hard". 5 µL of water were added stepwise until 40 µL volume was reached. The volume of the droplet, the length of the contact line, and the number of lamellae that were in contact with the water drop were analyzed during the measurements of advancing and receding contact angles. First experiments showed that the wetting properties of fabricated surfaces depend on the temperature in good correlation to the preliminary test in **Figure 3**. The lamellae remain undeformed upon the increase of the size of the droplet when they are in the hard state (E = 50 MPa, treated by cooling to 4 °C). On the other hand, lamellae in soft states (E = 9 MPa, T = 70 °C) were deformed by an advancing and receding water droplet. Interestingly, the contact angle undergoes irregular changes upon increase/decrease of its size that is due to jumping of the droplet between neighboring lamellae. It is clear – wetting properties depend on the mechanical properties of lamellae, which in turn depend on the temperature.

The deformation amplitude of lamellae is determined by a balance of surface tension ($F_s \sim \gamma \cdot dl$),[77] elastic deformation ($\delta \sim F \cdot H^3/EI$, where $H$ is the height of lamella, $I$ is second moment of inertia $I = a \cdot dl^3/12$, $a$ is the thickness of a lamella, $dl$ is its length), and gravity ($F_g \sim \rho \cdot g \cdot h \cdot b \cdot dl$) where $h$ is the height of the droplet, S – surface area (**Figure 4a**). We used the following values $H = 10^{-3}$ m, $a = 3 \cdot 10^{-5}$ m, $\delta = H/2 = 5 \cdot 10^{-4}$ m, $h = 1 \cdot 10^{-3}$ m for estimation of linear density of the surface tension forces and gravity forces which are $F_s/dl \sim 72 \ mN$ and $F_g/dl \sim 20 \ mN$. This means that gravity plays a considerable role and cannot be completely excluded from consideration. For simplicity, we however neglect effect of the gravitiy and the amplitude of lamellae deflection is determined by surface tension. The deflection is $\delta \sim \frac{\gamma \cdot H^3}{E \cdot a \cdot dl^2}$ and it increases linearly with the decrease of elastic modulus. This equation was derived for the case when a droplet is large or lamellas are small and whole lamella is deformed. In our case, only a part of lamella is deformed, therefore, the equation can be used only for quantitative estimation of dependences. An almost sixfold decrease in the elastic modulus from 50 MPa to 9 MPa (**Table S1**) leads to the same increase in deformation amplitude. Further, it increases with height (H) to the power of three. A threefold increase in the height (from 0.5 mm to 1.5 mm) leads to the 27-fold increase in deformation. Indeed, the low aspect ratio is the reason why structures prepared by photolithography and molding techniques[28, 35, 78] were not deformed by droplets and manual deformation is required to program their shape. The main advantage of our methods is that we are able to fabricate high aspect ratio structures – we were able to achieve an aspect ratio of 75 (larger aspect ratios are also possible), which either can or cannot be deformed by a droplet depending on their mechanical properties.

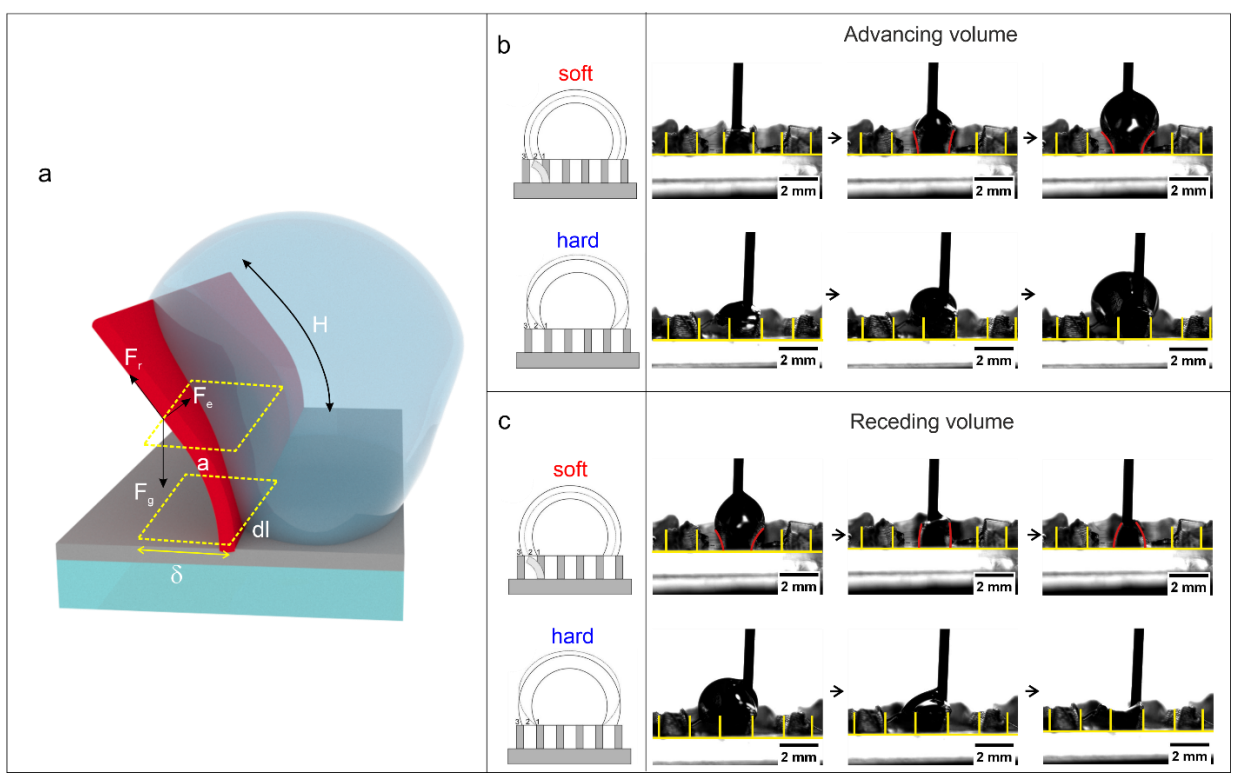

**Figure 4.** Deformation behavior of lamellae made from TPU PBA-75 (a) as influenced by advancing (b) and receding (c) water volume in the elastic (soft) and rigid (hard) state. Lamellae have been pushed or pulled away according to the direction of the volume of the drop.

The deformation of lamellae was quantified by measuring the projected length of the contact line, which is also referred to as "contact length" (**Figure S4**). We also tracked the number of lamellae that were in contact with the liquid. Since the distance between lamellae is known (it is determined during fabrication), we are able to correlate the contact length to the number of lamellae as expressed in **Figure 5** to understand the degree and direction of lamellae deformation. In the case when lamellae are undeformed, the contact line shall be equal to interlamellar spacing (d) multiplied by the number of the spacings (n) (which is n = $N$ -1 with N – the number of lamellae): $L_{contact} = d \cdot (N - 1)$. The experimentally measured contact length can be different from the theoretically estimated one when lamellae are deformed. The bending of the lamellae in the direction away from the droplet should lead to a higher value of the experimentally measured contact length. This is typically observed during advancing contact angle measurements (**Figure 4b**). The contact length (solid dots) is higher than the number of lamellae (empty dots) in **Figure 5**. The bending of the lamellae in the direction to the droplet (receding volume) should lead to a lower value of the experimentally measured contact length. Indeed, we observed that at a small interlamellar distance (500 µm) the contact length measured on soft surfaces is larger than that on hard surfaces during advancing and receding volume. Moreover, the measured contact length is larger than the calculated one during droplet advancing and is smaller during droplet receding. As a result, the advancing contact angle on soft surfaces is smaller than that on hard surfaces (**Figure S5**). As the distance between the lamellae increased, more volume was required to deform the same number of lamellae (**Figure 5**). For samples with distances of 1500 µm (**Figure S5**), part of the volume added dissipates into the groove, while the other part is inflating for the deformation of the lamellae, which accordingly is less volume than for lamellae with a shorter distance. Thus a lower number of lamellae is deformed. In comparison, small dimensions of the topographic surface, as occurred with the sample 500 µm in height and 500 µm in distance is deformed with less volume (6 to 10 µL) of the water drop. Results obtained at a large interlamellar distance (1500 µm) are difficult to interpret because in this case, the size of the droplet is comparable to the distance between lamellae and the value of contact length if the droplet was larger enough to be able to touch the next lamella. Droplets of nearly the same size, one of which is able to touch lamella and one, which is not able to touch lamella, show completely different wetting behavior.

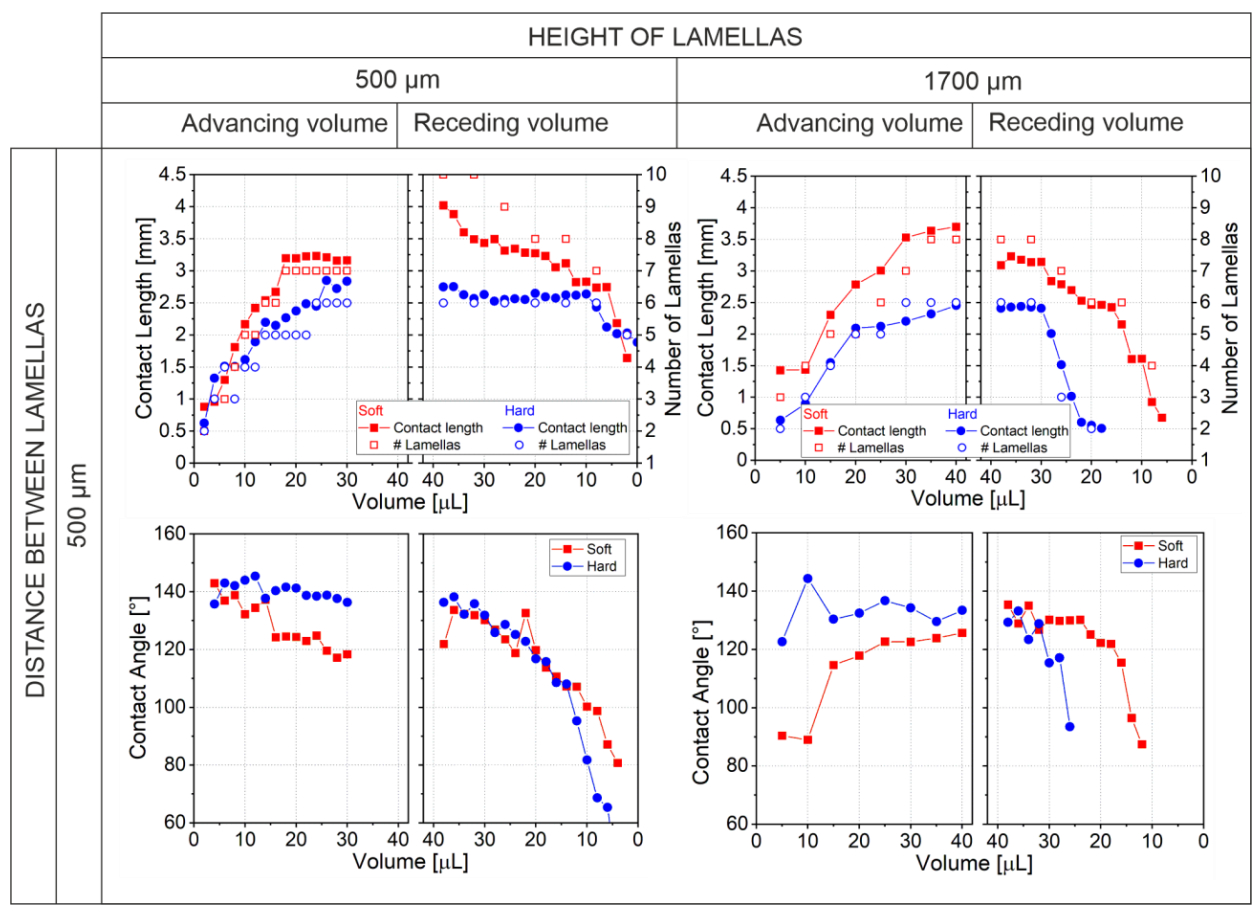

**Figure 5.** Effect of the volume of a water droplet on the contact length between the droplet and the surface, and the number of lamellae in contact with the water droplet. The analysis has been made for both states of the lamellae, 'soft' and 'hard'.

We performed a detailed investigation of the evolution of contact angle on hard and soft surfaces when the droplet size increases (**Figure 6**). It was found that contact angles fluctuate during the deformation of the lamellae. This fluctuation is due to the fact that jumping of the water drop to the next lamella requires certain activation energy. The contact angles increase upon inflation of droplets and drops down when it jumps to the next lamella. Generally, the contact angles on both hard and soft surfaces decrease upon the increase of the size of the droplet. The decrease on soft surfaces is however more pronounced. We explain this observation by the fact that lamellae in the soft state are more easily deformable and this deformation decreases contact angle as well as allows easier jumping to the next lamella. On contrary, droplets on the hard surface require stronger inflation to be able to jump to the next lamella. As result, contact angles on the soft surface are always smaller than on hard ones, and this difference increases with the distance between lamellae and their height. These observations clearly show that elastic force dominates over surface tension when lamellae are hard, while the deformation of soft lamellae is a result of the balance of surface tension and elastic force.

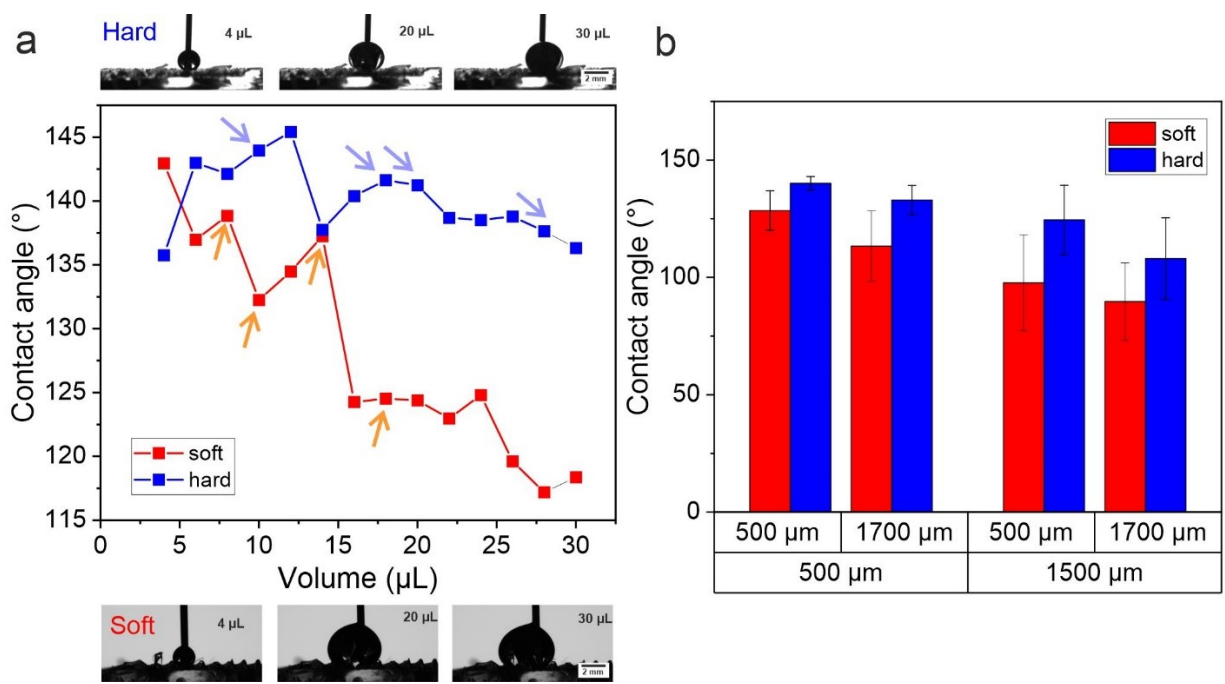

**Figure 6.** Advancing contact angle when the TPU PBA-75 is in a soft and hard state. a) example of the advancing angle for small topographies (500 μm height and 500 μm distance). Arrows show when the drop jumps to the next lamella (purple: hard state, orange, soft state). b) Average advanced angle for the different topographies constructed.

The TPU PBA-75 which was used for the fabrication of surfaces exhibits shape-memory behavior. To prove this, we programmed the shape of the lamellae. Therefore, a water droplet was placed on a structured surface at 70 °C (**Figure 7a, d**). As a result, the droplet deformed the lamellae as discussed above. Subsequently, the surface, with the droplet sitting on deformed lamellae, was cooled down to 4 °C and then the droplet was removed (**Figure 7b, e**). The lamellae stayed in a deformed state even at room temperature because the melt transition temperature of the PBA soft segment was not exceeded. Heating to 70 °C results in restoration of the shape of lamellae due to the melting of the soft segment and the associated triggering of the shape-memory effect (SME) (**Figure 7c, f**). The experiment was repeated more than 5 times, demonstrating the reliable shape-memory properties of the lamellae.

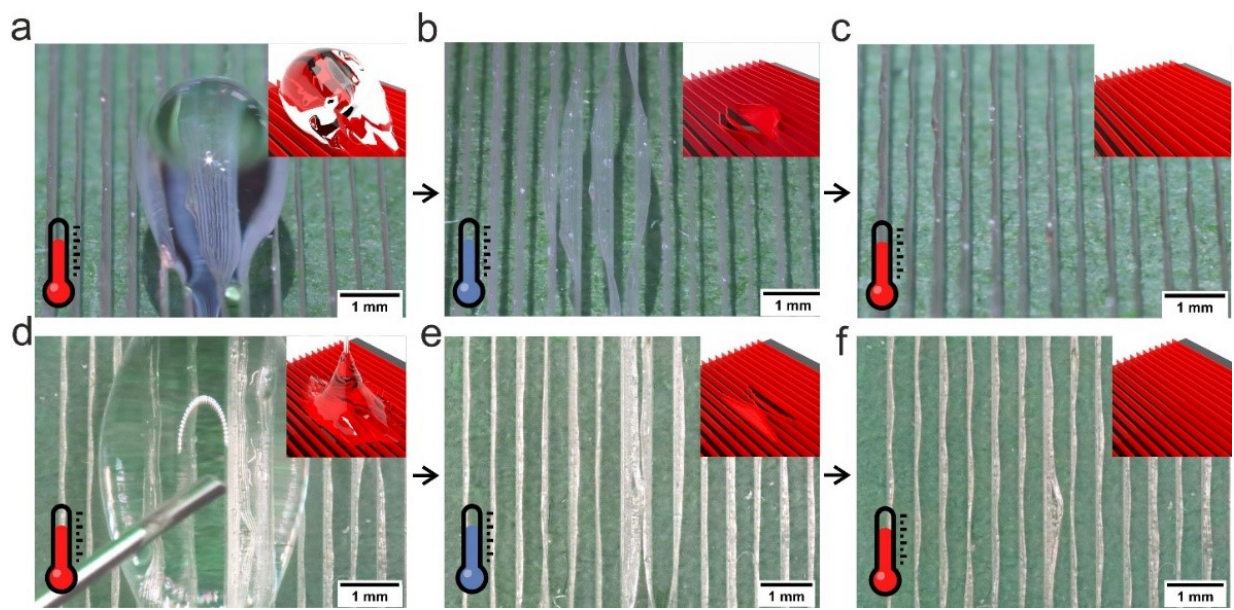

**Figure 7.** Shape-memory behavior of TPU PBA-75 when used as lamellae structured surface: a,d) water droplet-induced deformation of the lamellae at 70 °C, b,e) fixed shapes at 4 °C after removal of the drop, and c,f) the recovery of the initial surface by heating to 70 °C. This study was made for advancing the volume of the water droplet (a,b,c) and also for the receding volume (d,e,f).

Finally, we demonstrated the technological applicability of our insights by the combination of temperature-controlled deformability by droplets as well as SME and memory of mechanical properties for the fabrication of a smart valve, which can be controlled/programmed by a single water droplet (**Figure 8**). For this purpose, we fabricated lamellae with a distance of 1000 μm. The valve was made by making cuts in the direction perpendicular to one of the lamellae that formed a section, which can be bent (**Figure 8a**). The shape and deformability of this section can be programmed by temperature. When initially exposing the system to 4 °C and storing it at that temperature or room temperature, the valve is hard and cannot be deformed by a droplet deposited in the channel formed by lamellae (**Figure 8g**). The "smart valve" can be opened by advancing droplet at elevated temperatures (T = 70 °C) (**Figure 8b**). It can be fixed in a (i) open state, when it is bent and water is able to flow to the next channel. But the valve can remain in (ii) close state, when the lamellae has not been deformed by advancing volume, containing the fluid inside of one groove (**Figure 8g**). Switching back to the initial state (closing the valve) is possible by heating to 70 °C in order to trigger the SME (**Figure 8f**).

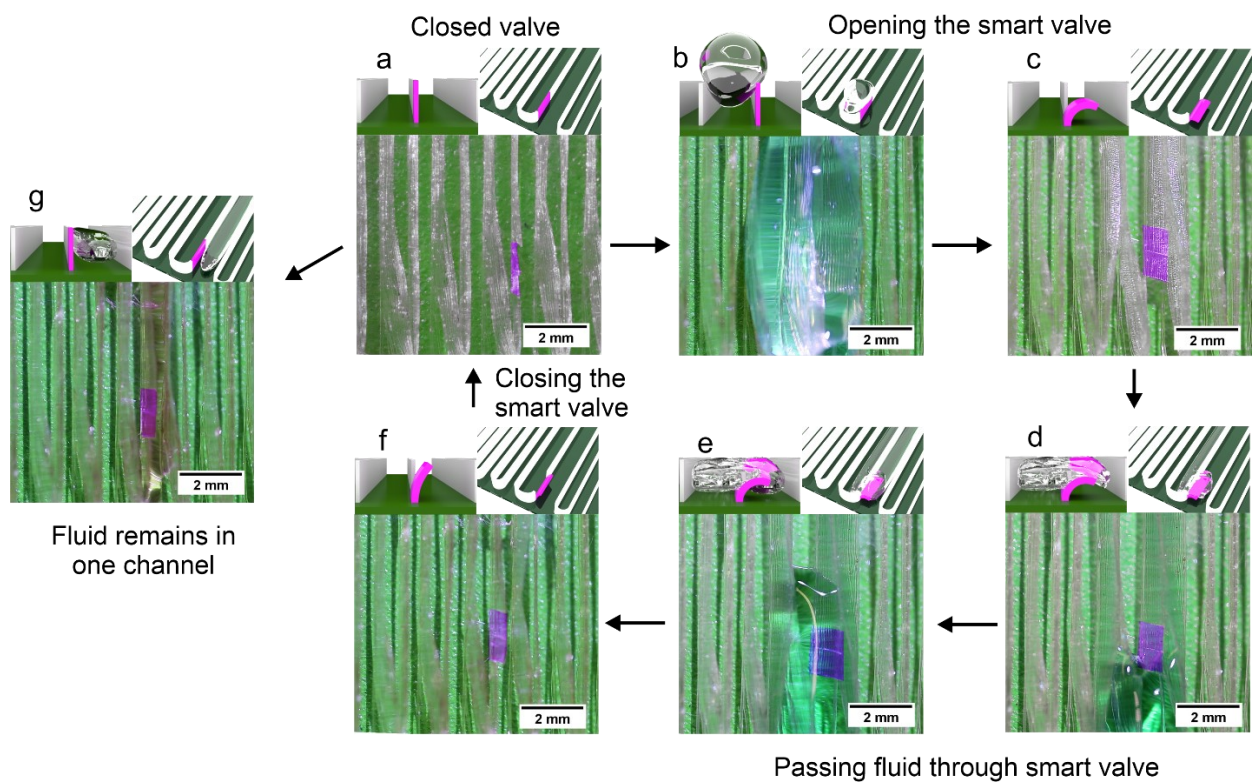

**Figure 8.** Application of the topographic surface of TPU PBA-75 for directing the flowing of a liquid by creating a smart valve: a) the smart valve (purple) is made by scission of one lamella, b) advancing volume of water drop for deforming the lamellae (T=70 °C) and then fixating at 4 °C, c) the smart valve bent (open state), d & e) the smart valve remained open at room temperature while the fluid passed through, f) the closing of the intelligent valve is carried out by increasing the temperature to 70 °C, and g) the fluid can not be directed to the next groove due to the closed smart valve.

## 3. Conclusion

This article reports for the first time on the fabrication and investigation of wetting properties of structured surfaces with an extraordinary aspect ratio of features (height/width ratio was 57:1, even a height/width ratio up to 75:1 was possible) as realized by using a polymer with promising thermomechanical and shape-memory properties. It was found that wetting properties of structured surfaces depend on temperature: droplets jump from one lamella to the next one at elevated temperature when the polymer is in a soft state resulting in a lower contact angle. The lamellae get deformed by droplets at elevated temperatures and the deformed state can be temporarily fixed by cooling to 4 °C. The temporary topology recovers to the original one as soon as the structure is heated above the switching temperature of the shape-memory polymer. Thus, the high aspect ratio allows tuning of geometry not only manually, as it is done in most works reported previously, but can also be done by placing a liquid, in this case a water droplet. Further, it is controlled by temperature - a liquid in combination with certain temperature conditions leads to distinct topography and wetting properties. This opens new opportunities for the design of smart elements of microfluidic devices such as, valves, which (i) cannot be opened at low temperature, (ii) can be opened by liquid at high temperature, (iii) can be left in an open state after cooling when liquid is applied and (iv) close at elevated temperature if no liquid is applied.

## 4. Experimental Section

*Materials:* The topographical surface was constructed with the thermoplastic polyurethane TPU PBA-75, characterized by a soft segment content of 75 weight percent. For the synthesis, Desmophen® 2505, which is a poly(1,4-butylene adipate) (PBA) diol was supplied by Covestro Deutschland AG (Leverkusen, Germany), 4,4´-diphenylmethane diisocyanate (MDI) was purchased from Fisher Scientific (Schwerte, Germany) and 1,4-butanediol (BD) as well as molecular sieve with a pore size of 4 Å were obtained from Alfa Aesar (Kandel, Germany). Microscope glass slides were coated with a solution of polylactide (PLA 4032D) from NatureWorks Ltd. (Minnetonka, MN, USA) and dichloromethane from Merck (Darmstadt, Germany).

*Synthesis of TPU:* TPU PBA-75 was synthesized by the prepolymer method as recently reported.[54] For this purpose, 0.037 mol of the PBA-diol Desmophen® 2505 was melted and dried overnight at 80 °C in a vacuum oven and 0.120 mol BD was stored over molecular sieve at ambient temperature. The next day, the PBA-diol was heated to 120 °C under nitrogen flow and stirring. 0.157 mol of molten MDI was added and the polymer melt was stirred continuously for about 90 min. Subsequently, the synthesized isocyanate-endcapped prepolymer was converted into TPU PBA-75 by adding BD as a chain extender. The stirring speed was then increased, and the viscosity change was monitored with an IKA® Eurostar 60 control from IKA-Werke GmbH & Co. KG (Staufen, Germany). Once the viscosity increased significantly, the reaction was stopped by pouring the melt onto a plate covered with a polytetrafluoroethylene film. Finally, TPU PBA-75 obtained was cured in an oven at 80 °C for 120 min, before being ground into granules for MEW.

*Spin-coating of the glass slides.* Microscope glass slides were spin-coated with a solution of PLA in dichloromethane (10 mg mL$^{-1}$). The rotational speed was set at 500 rpm for 10 s to ensure total coverage of the surface, and then, it increased to 4000 rpm for 1 minute.

*Fabrication of the surfaces.* The topographical structure was made via melt-electrowriting of TPU PBA-75 in a 3D Discovery printer Regen Hu (Villaz-St-Pierre, Switzerland). Parameters as applied voltage, pressure, movement rate, and temperature were adjusted for the fabrication of microfibers during melt-electrowriting. The TPU was melted at 215 °C. The voltage varied between 3 and 5 kV, and the distance set between the needle and the collector was set to 2 mm. Pneumatic pressure was used for the extrusion of the molten material and was fixed at 0.025, 0.05, 0.1, and 0.2 MPa. The movement rate (F) of the printhead was 10 mm s$^{-1}$. A metallic needle of 200 μm inner diameter was used to extrude the molten polymer. The lamellae were fabricated with the addition of linear fibers piled up one over the other. Height and distance between the lamellae varied between 500, 1250, and 1700 μm height and distances of 500, 1000, and 1500 μm.

*Scanning Electron Microscopy (SEM).* A scanning electron microscope (SEM) Thermo Fischer Scientific Apreo 2 SEM (Germany) was used for measuring the diameter and analyzing the morphology of MEW fibers. The samples were fixed on SEM stubs using cupper adhesive tape and covered with ~1.3 nm platinum to ensure conductivity using a Leica EM ACE600 (Wetzlar,

Germany). The sputtering rate was set at 0.02 nm s$^{-1}$, with a current of 35 mA, under Argon 0.05 mbar.

*Differential Scanning Calorimetry (DSC).* DSC 3 Mettler Toledo (Greifensee, Switzerland) was used for the thermal analysis of the TPU PBA-75. It was performed two heating curves and one cooling curve. The first heating was conducted to erase the thermal history of the polymer. The heating rate was set at 1 °C min$^{-1}$ under nitrogen flow.

*Dynamical Mechanical Analysis (DMA).* A Modular Compact Rheometer MCR 702 Multidrive from Anton Paar GmbH (Ostfildern, Germany) was used to perform the DMA analysis for the mechanical properties of the polymer, also cyclic extension measurements were made to analyze the elasticity of the material. A TPU filament with a diameter of 1.4 mm was used to determine the mechanical properties at temperatures between 20 and 100 °C. Due to limitations in the cooling system, the material was exposed to 4 °C overnight prior to the measurements, to investigate the effect of the crystallization on the material. The frequency was kept at 1 Hz.

Based on the cyclic experiments, it was calculated the recovery and resilience percentage of the material. Following the same protocol as mentioned in our publication,[79] the recovery was calculated by the relation between the residual strain after the unloading step ($\varepsilon$) and the original strain ($\varepsilon_0$) ($\% recovery = (1 - \varepsilon/\varepsilon_0) \cdot 100\%$). The resilience was obtained by the equation: $\% \, resilience = (A_{unload}/A_{load}) \cdot 100\%$, where $A_{unload}$ and $A_{load}$ represented the area under the stress-strain curve for the unloading and loading curve, respectively.

*Rheology.* The rheological properties of the molten TPU PBA-75 were evaluated by using the MCR 702 Multidrive Anton Paar (Ostfildern, Germany). Parallel plate geometry of 25 mm diameter was used to measure the viscosity at 200 °C. The angular frequency was varied from 0.1 – 100 Hz. While for the structured lamellae, a temperature ramp was performed from -5 to 20 °C and from 80 °C to 20 °C, to analyze the temperature dedendence of mechanical properties. The sample was fixed to the base and with a parallel plate geometry of 25 mm diameter, it was measured the storage and loss modulus at a constant frequency of 1 Hz and constant shear strain of 1 %.

*Drop Shape Analysis.* A Krüss Drop Shape Analyzer DSA25 (Hamburg, Germany) was used to determine the variation of the contact angle. The sessile drop method was used to determine the contact angle of the flat surface composed of spin-coated PLA. For advancing and receding volume studies, a total of 40 µL of ultrapure water was increased or decreased in multiple doses of 5 µL at a rate of 2.67 µL s$^{-1}$. The measurements were made at room temperature.

The shape-memory behavior of the structured topographical surface was studied by deforming the lamellae at 70 °C. By advancing and receding the volume of a water drop, the lamellae were able to deform in direction of the water flow. The sample was fixed by cooling to 4 °C. The recovery of the original state of the sample was accomplished by heating the sample to 70 °C.

**Supporting Information**

Supporting Information is available from the Wiley Online Library or from the author.

**Acknowledgments**

This work was supported by Deutsche Forschungsgemeinschaft (DFG) (Grants: IO 68/15-1, AL1705/5-1) and also by Fraunhofer Cluster of Excellence "Programmable Materials" under project 630507. T.P. wishes to thank the European Regional Development Fund for financing a large part of the laboratory equipment (project 85007031). G.C. & L.I. would like to acknowledge SPP 2171 for allowing to establish collaborations between investigation groups.

The manuscript was written through the contributions of all authors. All authors have given approval to the final version of the manuscript.

**Smart mechanically tunable surfaces with shape memory behavior and wetting-programmable topography**

ToC figure

This work reports how to control the deformation behavior of lamellae structured surfaces formed by a shape-memory polymer. The lamellae are characterized by a high aspect ratio and thermally adjustable mechanical properties with regard to their state at room temperature. The lamellae structure exhibits a wetting programmable topography.

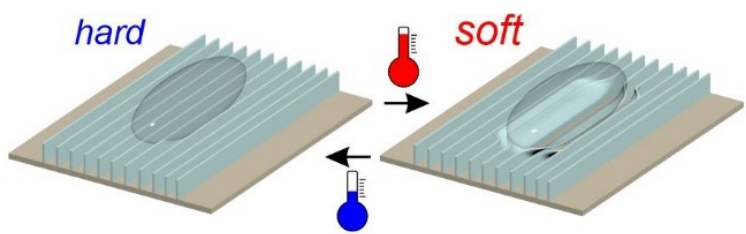